\documentstyle[12pt]{article}
\begin{document}
\rightline{TUHE-9641}
\begin{center}
{\bf\large Application of sum rules to heavy baryon masses}\\
\vspace{16pt}
Jerrold Franklin\\
{\it Department of Physics,Temple University,\\
Philadelphia, Pennsylvania 19022}\\
April, 1996
\end{center}
\begin{abstract}

Model independent sum rules are applied to recent measurements of
heavy c-baryon and b-baryon masses.  The sum rules are generally
satisfied to the same degree as for the light (u,d,s) baryons.

\end{abstract}
\vspace{.5in}

     Recent measurements of the masses
of a number of heavy baryons (with c or b quarks) now make it
possible to test model independent sum rules that were derived some
time ago using fairly minimal assumptions within the quark
model.\cite{cb} The sum rules depend on standard quark model
assumptions, and the additional assumption that the interaction
energy of a pair of quarks in a particular spin state does not
depend on which baryon the pair of quarks is in ("baryon
independence").
This is a weaker assumption than full SU(3) symmetry of the wave
function, which would require each individual wave function to be
SU(3) symmetrized.  Instead, we use wave functions with no SU(3)
symmetry as described in Ref.\cite{sqm}.
No assumptions are made about the type of potential, and no
internal symmetry
beyond baryon independence is assumed.
The sum rules allow any amount of breaking in the interactions and
individual wave functions, but do rest on baryon independence for
each quark-quark interaction energy.   A more detailed discussion
of the derivation of the sum rules is given in Ref.\cite{cb} In a
previous paper, we applied an isospin breaking sum rule to the
$\Sigma_c $ charge states.\cite{cb2}  In the present paper we test
sum rules that connect baryon states of different isospin, which we
characterize as medium strong energy difference sum rules.

     Before looking at the heavy baryon masses, we review the
application of the model independent sum rules to the light
baryons.  There, the following sum rules have been
derived\cite{sqm}
\vskip .3in
$\begin{array}{cccccr}
\frac{1}{3}(\Omega^{*-}-\Delta^{++})
      & = & \Xi^{*0}-\Sigma^{*+} & = &
\Xi^0-\Sigma^+,&\hspace{1.5in}(1)
\\ [.1in]
       (147\pm 1) && (149\pm 1)&& (125)
\end{array}$
\vskip .3in
$\begin{array}{cccr}
2\overline{N}+2\overline{\Xi}-3\Lambda-\overline{\Sigma} &=&
\Omega^* +\overline{\Delta}
-\overline{\Sigma^*}-\overline{\Xi^*}. & \hspace{1.6in}(2)\\
(-26)& & (-14\pm 2)
\end{array}$
\vskip .3in
\noindent
The baryon symbol has been used for its mass, and, except for the
$\Delta$, a star indicates  spin $\frac{3}{2}$.  A bar over the
symbol represents an average over the particular isospin multiplet.

Where specific charges are indicated, these could be changed using
the isospin breaking sum rules in Ref.\cite{cb}.
The experimental value in MeV for each sum is given below each
equation.

     The Gell-Mann Okubo formula and equal spacing in the decuplet,
which would follow if SU(3) symmetry were broken only by a small
octet component, correspond to each side of Eq.\ (2) being zero.
The deviations of each side in Eq.\ (2) from zero indicate that the
light baryon interactions
do not satisfy that assumption (which is not required for the sum
rules), with breaking of the order of 10 MeV.
The deviation between the two sides in Eq. (2) and by the
$\Xi^0-\Sigma^+$ term in Eq.\ (1) indicates that the light baryon
wave functions also violate our baryon independence assumption to
the extent of about 10-20 MeV in the masses.  We note that the
breaking of baryon independence occurs for the two cases where
baryons of different spin are compared.  Similar breaking should be
expected in the heavy baryon sum rules, even if the heavy spin-spin
interactions are more SU(3) and SU(6) symmetric.

     For extension to heavy baryons, it is convenient to use the
equalities in Eq.\ (1) to replace Eq.\ (2) by
\vskip .3in
$\begin{array}{cccccr}
(\Delta^+ -p) &=& (\Sigma^{*0}-\Sigma^0)+\frac{3}{2}(\Sigma^0
-\Lambda^0)
&=& (\overline{\Xi^*}-\overline{\Xi}) +\frac{3}{2}(\Sigma^0
-\Lambda^0). & (3)\\ [.1in]
(297) & & (307) & & (330) &
\end{array}$
\vskip .3in
\noindent
Equation (3) shows a spread of $\sim$ 30 MeV corresponding to
breaking of that order of baryon independence in the light baryons.
This demonstrates the ambiguity that arises when the sum rules are
extended to heavy baryons.  We will try to minimize this ambiguity
by choosing the most suitable light baryon combination to compare
to heavy baryons.

     In table I, we list the measured heavy baryon masses that will
be used in the sum rules.
\begin{table}
\centering
\begin{tabular}{llc}
Baryon & Mass (MeV) & Reference\\
\hline\hline
$\Sigma_c^{++}$ & $\Lambda_c^+ +168.0\pm 0.3 $&\cite{pdg}\\
$\Sigma_c^{+}$ & $\Lambda_c^+ +168.7\pm 0.4 $&\cite{pdg}\\
$\Sigma_c^{*++}$ & $\Lambda_c^+ + 245\pm 7 $&\cite{skat}\\
$\Xi_c^{\prime+}$ & 2563$\pm$15 &\cite{wa}\\
$\Xi_c^{*0}$ & 2643.3$\pm$2.2 &\cite{cleo}\\
$\Omega_c^0$ & 2700$\pm$3 & \cite{frab}\\
$\Sigma_b^-$ & $\Lambda_b^0 +173\pm 9$ & \cite{delphi}\\
$\Sigma_b^{*-}$ & $\Lambda_b^0 +229\pm 6$ & \cite{delphi}\\
\hline
\end{tabular}
\caption{Heavy baryon masses used in the sum rules.}
\end{table}
We indicate the expected baryon assignments in table I. The
$\Xi_c^{\prime+}$ is the spin $\frac{1}{2}$ usc baryon having the
u-s  quarks in a spin 1 state.
We extend Eq.\ (3) to charmed baryons by changing the s-quark into
a c-quark.  This leads to\cite{comb}
\vskip .3in
$\begin{array}{cccr}
(\Sigma^{*0}-\Lambda^0)+\frac{1}{2}(\Sigma^0 -\Lambda^0)
&=&(\Sigma_c^{*+}-\Lambda_c^+)+\frac{1}{2}(\Sigma_c^+ -\Lambda_c^+)
& \hspace{.8in}(4)\\ [.1in]
(307) & & (330\pm 7) &
\end{array}$
\vskip .3in
\noindent
We have used the light Sigma baryons for the left hand side of Eq.\
(4).  Use of other combinations could change the left hand side, as
indicated in Eq. (3), but the Sigma combinaton is the most
reasonable since they are most similar to their charmed
counterparts.
Equation (4) is written in terms of differences from the $\Lambda$
mass which is how the $\Sigma_c$ and $\Sigma_c^*$ masses are
measured.  We have had to use the
measured $\Sigma_c^{*++}$ mass for the $\Sigma_c^{*+}$ mass in
Eq.(4), but
that difference is probably small.

  Changing the c-quark in any c-baryon sum rule to a b-quark leads
to the corresponding sum rule for b-baryons.
Applying this to Eq.\ (4) leads to
\vskip .3in
$\begin{array}{cccr}
(\Sigma^{*0}-\Lambda^0)+\frac{1}{2}(\Sigma^0 -\Lambda^0)
&=&(\Sigma_b^{*0}-\Lambda_b^0)+\frac{1}{2}(\Sigma_b^0 -\Lambda_b^0)
  &\hspace{1in}(5)\\ [.1in]
(307) & & (316\pm 10) &
\end{array}$
\vskip .3in
\noindent
The sum rules in Eqs. (4) and (5) are satisfied to about the same
extent as the light baryon sum rules.

     In Ref.\cite{cb2} we used a sum rule to predict the
$\Xi_c^{\prime}$ mass which has now been measured.  This permits a
test of the sum rule, which we write here as
\vskip .3in
$\begin{array}{cccr}
\Sigma^+ +\Omega^{*-} -\Xi^0 -\Xi^{*0} &=& \Sigma_c^{++}+
\Omega_c^{0} -2\Xi_c^{\prime +}. & \hspace{1.5in}(6)\\ [.1in]
(15) & & (27\pm 30) &
\end{array}$
\vskip .3in
\noindent
Again, we have used the combination of light baryon masses most
similar to the corresponding charmed baryons.  However, the left
hand side of Eq.\ (6) could be made to vary between -3 and +27 MeV,
by using
Eq.\ (1) to substitute other light baryon combinations .

     The spin $\frac{3}{2}$ counterpart of Eq.\ (6) can be used to
predict the as yet unmeasured $\Omega_c^{*0}$ mass
\setcounter{equation}{6}
\begin{equation}
\Omega_c^{*0}=\Omega_c^{0}+2(\Xi_c^{*+}-\Xi_c^{\prime+})
-(\Sigma_c^{*++}-\Sigma_c^{++})=2783\pm30,
\end{equation}
where we have used the most similar c-baryon combination rather
than using any light baryons.  This increases the error on the
prediction, but is the more reasonable choice.

     We see that, especially when the most reasonable combination
of  light baryons is taken, the medium strong energy difference sum
rules are satisfied at least as well for the heavy baryons as for
the light quark baryons.  However the situation is not as nice for
the isospin breaking mass differences in the case of the
$\Sigma_c$.  In Ref.\cite{cb2} we showed that the $\Sigma_c$
sum rule is violated by three standard deviations, while the
corresponding light baryon sum rule is satisfied.  Since sum rules
in disagreement are of more concern than those which are satisfied,
resolving the $\Sigma_c$  mass differences is of prime importance.

     I would like to than Don Lichtenberg for useful comments about
this work.


\begin{thebibliography}{8}
\bibitem{comb}Equation (4) is a combination of several equations in
Ref. [1].
\end{thebibliography}
\end{document}